# An Efficient Coding Method for Spike Camera using Inter-Spike Intervals

Siwei Dong[*], Lin Zhu[*,1], Daoyuan Xu[+], Yonghong Tian[*,2] and Tiejun Huang[*]

*School of EECS,  
Peking University  
Beijing, P.R. China

[+]School of ECE, Shenzhen Graduate School,  
Peking University  
Shenzhen, Guangdong, P.R. China

*Abstract*: Recently, a novel bio-inspired spike camera has been proposed, which continuously accumulates luminance intensity and fires spikes while the dispatch threshold is reached. Compared to the conventional frame-based cameras and the emerging dynamic vision sensors, the spike camera has shown great advantages in capturing fast-moving scene in a frame-free manner with full texture reconstruction capabilities. However, it is difficult to transmit or store the large amount of spike data. To address this problem, we first investigate the spatiotemporal distribution of inter-spike intervals and propose an intensity-based measurement of spike train distance. Then, we design an efficient spike coding method, which integrates the techniques of adaptive temporal partitioning, intra-/inter-pixel prediction, quantization and entropy coding into a unified lossy coding framework. Finally, we construct a PKU-Spike dataset captured by the spike camera to evaluate the compression performance. The experimental results on the dataset demonstrate that the proposed approach is effective in compressing such spike data while maintaining the fidelity.

## 1. Introduction

In recent years, bio-inspired vision sensors have become very attractive in the field of self-driving cars, unmanned aerial vehicles and autonomous mobile robots, due to their significant advantages over conventional frame-based cameras, such as high dynamic range and fast sensing capability. It is accustomed to name the output data of these sensors as spikes in a similar manner to biological systems. Typically, the pixels, also known as the photoreceptors, independently respond to the luminance intensity changes with spikes elicited. In contrast to frame-based cameras, bio-inspired vision sensors all have in common is frame-free. From this perspective, the temporal resolution can be as high as possible, because there is no need of shutters. This is reasonable if we turn attention to biological retina systems, in which the photoreceptors independently convent incoming light into electrical signals that are transmitted to the inner plexiform layer of the retina and further processed as spikes conveyed to the visual cortex via the optic nerve. The independent sampling breaks the limitation of exposure which is commonly the bottleneck for most frame-based cameras to increase frame rates.

To simulate biological vision, one of the most famous artificial "silicon retinas" is the dynamic vision sensor (DVS) [1]. It is capable of high speed detection and tracking. However, as it only cares about the relative change of luminance intensity, it is very difficult to reconstruct the texture. To improve this, Brandli *et al*. [2] proposed the dynamic and active-pixel vision sensor (DAVIS) which integrates the DVS with the frame-based active-pixel sensor (APS) together. The solution makes it able to capture both frames and DVS spikes, but the mismatch is quite obvious due to the lower frame rate of APS (60

---

frames per second) in contrast to the DVS with a temporal resolution of 1 $\mu s$. Posch *et al.* [3] tried a different way by introducing the asynchronous time-based image sensor (ATIS). ATIS consists of a DVS circuit and a photo-measurement circuit. Once a DVS spike is fired (i.e. intensity change detected), the photo-measurement is triggered and the intensity is acquired by encoding the time from the change detection to the threshold crossing of a photocurrent integrator. Since the intensity is inversely proportional to the integration time, it is able to reconstruct the visual texture of the scene. ATIS seems to be perfect, but the intensity is only measured when a DVS spike is generated, and the measurement is posterior to the spike firing, so the mismatch still exists, especially in flexible motion.

The key issue is the sampling mechanism of DVS which only responds to the relative intensity change. For stationary scenarios, there will be few spikes fired. Thus it is difficult to recover the absolute intensity. In our view, as a vision sensor, the ability to record the scene is indispensable. Aiming at this goal, we deeply investigated the sampling problem and finally designed the spike camera [4] which was enlightened from the integrate-and-fire neural model. The spike camera has a spatial resolution of 400×250 and the maximum spike firing rate per pixel is 40,000 Hz. In the spike camera, each pixel operates independently accumulates luminance intensity which inputs from an analog-to-digital converter (ADC), and generates a spike as soon as possible if the ADC value exceeds the dispatch threshold $\phi$:

$$\int_0^t I dt \geq \phi \tag{1}$$

where $I$ refers to the luminance intensity (usually measured by photocurrent in the circuit). Then the accumulator is reset and all the charges on it are drained. At different pixel, the accumulation speed of the luminance intensity is different. As shown in Fig.1, the brighter pixel will send out spikes more frequently than the darker one, because more photons are collected at the pixel, and its intensity accumulation speed is faster which is easier to exceed the threshold. Compared to conventional frame-based cameras and DVS, the spike camera has successively achieved the balance between both high speed motion capture and consistent texture reconstruction.

Although the output data are called spikes as well, the characteristics differ from those of DVS. According to DVS sampling mechanism, a spike represents the relative intensity change, therefore the temporal redundancy is almost removed. For example, if the intensity is stable, nothing will be outputted from DVS. But in the spike camera, both the static background and the moving foreground objects are firing spikes with various frequencies.

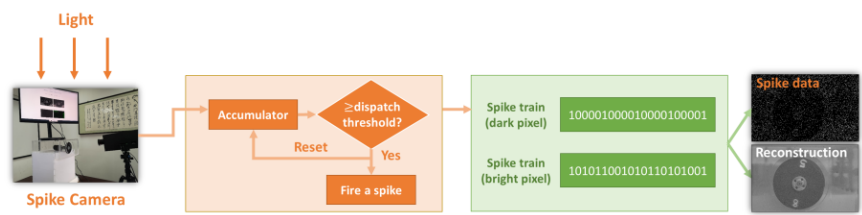

**Figure 1:** The workflow of the spike camera

Due to high temporal resolution and the spatial continuity of pixels, the spike data have redundancy in both temporal and spatial domain. According to the statistics, the spike camera generates about 476 megabytes of data per second. Thus, the compression of the spike data will be of benefit for storage and transmission. In this paper, we design an efficient coding framework to address this problem.

The rest of the paper is organized as follows: Section 2 briefly reviews related work, and Section 3 analyzes the distribution of the spike data. In Section 4, presents the spike coding framework. In Section 5, the experiments on compression performance and the distortion evaluation are demonstrated. Finally, the paper is concluded in Section 6.

## 2. Related Works

Spike coding of bio-inspired cameras is a new research field recently proposed. The spike data is usually with a very high temporal resolution, which consumes more storage space. The compression for the spike data is a challenge since its data characteristics are very different from conventional videos. In [5], Bi *et.al.* first explicitly raised the issue of DVS data compression, a lossless coding algorithm of spike data was proposed and achieved an impressive coding performance. This is a good reference for our work. However, the DVS data is quite different from our spike data. DVS sensor outputs spikes only when the luminance changes, so its data is much sparser than that of the spike camera and the data characteristics of both are completely different.

The spike camera is inspired by biological neuron model, this drives us to explore the coding technology from the neural computation field. Initial methods of applying entropies to neural coding treated spike trains as binary strings by quantizing the time axis [6]. Another category is interval methods, which focused on intervals between consecutive spikes rather than on spikes themselves. In [7], Bhumbra *et al.* proposed that for a constant level of activity, the coding capacity of a single input is equal to its repertoire of inter-spike intervals. From the perspective of the neuron, the interval methods are more convincing than the methods based on spike counts [8]. Some probability density function such as gamma and Gaussian were often used to apply the interval methods [9]. Generally, a simple Poisson integrate-and-fire model can be described as a gamma distribution and close fits to recorded spike data [10].

On the other hand, some coding techniques in conventional video coding field may also be referential for spike coding, including motion compensation, quantization and entropy coding in AVC/H.264 [11] and HEVC/H.265 [12]. These technologies give us great inspiration in designing our coding strategies.

## 3. Spike Data Analysis

In this section, we investigate the probability density distribution of inter-spike interval (ISI), and the spatial and temporal distribution of spike data.

### 3.1. The Probability Density Distribution of Inter-Spike Interval

The distribution of inter-spike interval can be derived from the photon arrival process which is usually assumed to be a homogeneous Poisson process. It is parameterized by a single scalar $\lambda$ which gives the mean rate with which events arrive. Each photon arrival event is completely independent from all the others. Therefore, the probability of $n$ photon arrival events is

$$\Pr(N(\delta) = n) = e^{-\lambda\delta}\frac{(\lambda\delta)^n}{n!} \tag{2}$$

where $N(\delta)$ refers to the number of photons arrived during a period of time $\delta$; $\lambda$ is the photon arrival rate. In a very short period of time, $\lambda$ is constant.

Since the accumulated intensity is determined by the number of photons arriving, we assume that the arrival of $\vartheta$ photons will reach the dispatch threshold $\phi$ and generate a spike. If the spike firing time is denoted as $t_i$, the inter-spike interval $x = t_i - t_{i-1}$. As $N(\delta)$ is a Poisson process, the time from the beginning to the occurrence of $n$ photons arriving is a gamma distribution [13]. Thus, the probability density distribution (PDF) is:

$$f(x, \beta, \alpha) = \frac{1}{\beta^\alpha \Gamma(\alpha)} x^{\alpha-1} e^{-x/\beta}, x > 0 \tag{3}$$

where $\alpha$ is the shape parameter and $\beta$ is the scale parameter in gamma distribution, here $\alpha = \vartheta$, $\beta = 1/\lambda$, and $\Gamma(\cdot)$ refers to the gamma function.

According to the mechanism of the spike camera, the dispatch threshold is predefined leading to the shape parameter $\alpha$ to be constant, but for different pixels, the luminance intensity may be slightly different with various scale parameters $\beta$. Fig. 2 shows the actual ISI values are well fitted by the gamma PDF.

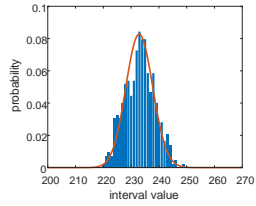
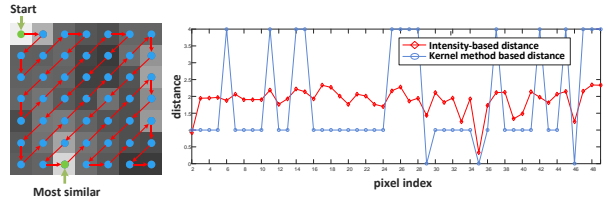

**Figure 2:** Histogram of the practical ISI values, and the fitted gamma probability density function (in red line).

**Figure 3:** The spatial distribution of spike data

Based on the above, we are able to analyze the temporal and spatial correlations of ISI. For a certain pixel, when the luminance intensity is stably changing or constant, the PDFs of ISIs between consecutive spikes tend to be similar. For the pixels in a spatially neighboring region, their photon arrival rates $\lambda$ have a great probability to be similar, and so does their PDFs. As a result, the ISI should have the temporal and spatial correlation.

### 3.2. The Spatial and Temporal Distribution of Spike Data

To begin exploring the spatial correlation of the spike data, we need to determine a measure to estimate the distance of two spike trains first. In our previous work [5], the kernel method is used to measure the spike train distance of DVS. However, it fails to accurately measure the spike trains from the spike camera. That's the motivation for us to explore a novel intensity-based distance. Considering that Eq. (1) can be simplified as $It \geq \phi$, where $t$ exactly corresponds to the ISI. Therefore, the average intensity of the pixel in this period can be estimated by

$$\bar{I} = \frac{\phi}{t} \tag{4}$$

For the visual system, the difference of intensities caused by the variation of ISI $\Delta t$ can be defined as

$$\Delta I = \frac{\phi}{t} - \frac{\phi}{t+\Delta t} = \frac{\phi \Delta t}{t^2 + t\Delta t} \tag{5}$$

Thus, for two spike trains $f_{s_1}$ and $f_{s_2}$, we can first convent them to two ISI sequences. Each ISI is denoted as $t_{s_1}^{(i)}$ and $t_{s_2}^{(i)}$, respectively. Here $i = 1, 2, .., K$ and $K$ refers to the number of ISIs. Then, the intensity-based distance between two spike trains is:

$$\|f_{s_1} - f_{s_2}\| = \sqrt{\frac{1}{K} \cdot \sum_{i=1}^{K} \left( \frac{1}{t_{s_1}^{(i)}} - \frac{1}{t_{s_2}^{(i)}} \right)^2} \tag{6}$$

Due to the high spatial correlation in actual objects, the photon arrival rate of spatially adjacent pixels is very similar, thus the spatial distribution of spike data should be also highly correlated. To explore the spatial distribution of the spike data, in the sequence "disk-pku", we select a $7 \times 7$ sized square area in a period of time. An arbitrary pixel is chosen as the start pixel, such as the one in the upper left corner. Then the distances between the start pixel and other pixels are respectively measured by Eq. (6), in a zigzag order. The distances are shown in Fig. 3, it can be seen that the spatially adjacent spike trains are with similar intensities, leading to a near distance. By applying the proposed measure, the most similar pixel can be accurately found out. In contrast, the kernel method gets three candidates but fails to figure out the right one.

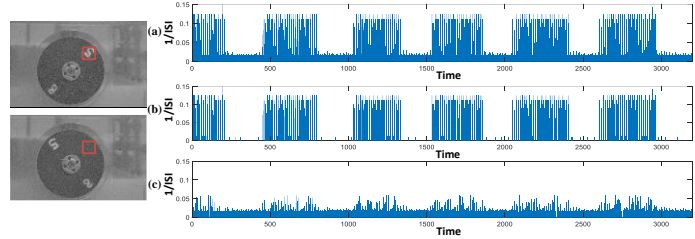

**Figure 4:** The temporal distribution of spike data. (a) The intensity (1/ISI) distribution in temporal domain; (b)The temporal distribution corresponding to bright object (the digits "5" and "8"); (c) The temporal distribution corresponding to dark object (the black disc).

As for the temporal correlation, the nature of continuous arrival of photon constitutes the temporal correlation between each consecutive spike of a certain pixel. In order to analyze the spike data intuitively, we give the distribution of the reciprocal of ISI (corresponds to the luminous intensity of each ISI) in the temporal domain. As shown in Fig. 4, the sequence "rotation" depicts a spinning disc at 2000 rpm. We select a pixel located in the red box (as shown in the left). When the pixel is dark, we can assume that the photon arrival rate $\lambda$ is constant, thus the PDF of ISI can be modeled as a gamma distribution. The digits "5" and "8" will be easily distinguished when they appear in the red box since they cannot fit the PDF. The results show clearly that the spike data has a high correlation in the temporal domain.

## 4. Spike Coding Framework

In this section, a spike coding framework is proposed to compress the spike data (Fig. 5). First, the spike train is adaptively partitioned into multiple segments in temporal domain. Then, the intra-pixel and inter-pixel coding including multiple prediction modes are performed to find the best reference candidate. Afterward, the prediction residuals are quantized to achieve lossy compression. Finally, the quantized residuals are fed into an adaptive context-based entropy coder. Overall, to achieve the best performance, each prediction mode will be tried and the best one with minimum rate-distortion cost is chosen.

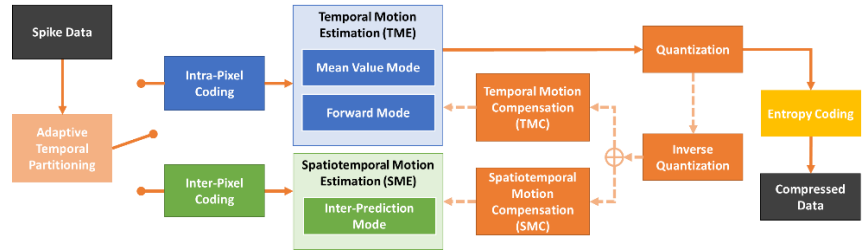

**Figure 5:** The framework of spike coding

## 4.1. Adaptive Temporal Partitioning

Generally, the basic coding unit is a block or a cube. However, the numbers of ISI at different pixels are quite different. In Fig. 6, even in a background region, the ISI numbers vary greatly from pixel to pixel. Consequently, in the proposed framework, the prediction and coding strategies are designed for pixels. For a certain pixel, considering the luminance intensity variation and the object's movements, the ISI may change significantly which is depicted in Fig. 6. In this case, it is reasonable that the ISIs of a pixel should be adaptively partitioned into multiple segments.

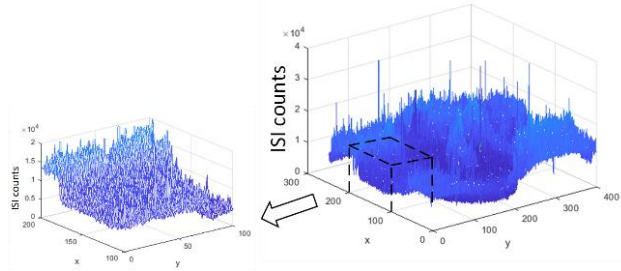

**Figure 6:** The distribution of ISI counts for all the pixels from the sequence "rotation".

Firstly, the whole ISI sequence is divided to basic segments (32 ISIs in our experiments). Since the distribution of ISI obeys gamma distribution in a short period of time, a basic segment can be well fitted by a gamma PDF, and the parameters $\alpha$ and $\beta$ are determined by the mean and variance of the ISIs. Then, by comparing the parameters $\alpha$ and $\beta$, we can determine whether two basic segments can be merged or not. The partitioning strategy can be iterated until all adjacent segments are in different distributions. Now the pixel contains multiple segments which can be well predicted utilizing temporal and spatial correlations.

## 4.2. Intra-pixel Coding

Since the pixels are independently respond to the luminance change, for real-time compression scenarios, the prediction and coding are limited in the segments from the same pixel, namely the intra-pixel coding. There are two prediction modes designed, mean value mode (MVM) and forward mode (FM) shown in Fig. 7.

*a) Mean value mode*

For a pixel in the background region, the ISIs are almost the same except for some noises. MVM is designed for this case. By subtracting the mean value of the ISIs within the segment, the residuals obtained are very close to zero which can be further quantized. The mean value is differentially coded between successive segments.

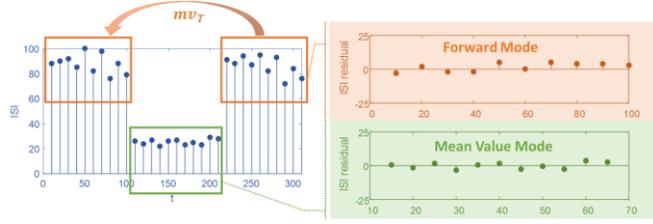

**Figure 7:** Mean value mode and forward mode in Intra-pixel coding.

*b) Forward mode*

In the sequence "rotation", the digits "5" and "8" appear periodically at a certain pixel. Due to fast moving, the ISIs within a segment may vary from each other which are not well predicted in MVM. FM utilizes temporal motion estimation (TME) and temporal motion compensation (TMC) to find a better reference candidate. For accurate prediction, the reference candidate is searched among all the previous coded ISIs. During TME process, Eq. (6) is used to measure the similarity. The temporal motion vector $mv_T$ is not coded directly. Instead, it is predicted by the motion vectors of previous $M$ available segments. Available segments refer to the segments with the same prediction mode of FM. Then the motion vector difference $mvd_T$ is coded.

$$mvPred_T = \frac{1}{M} \sum_{i=1}^{M} mv_{T_i} \quad \text{and} \quad mvd_T = mvPred_T - mv_T \tag{7}$$

### 4.3. Inter-pixel Coding

To cope with complex situations such as high speed motion, the inter-pixel coding is proposed by taking advantages of spatial correlations of the spike data. In inter-pixel coding, TME and TMC are enhanced to SME and SMC which enable the spatiotemporal search for the best reference candidate.

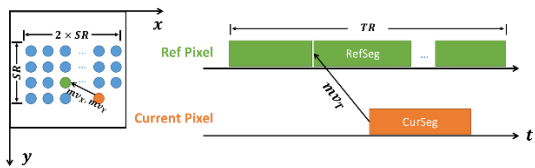
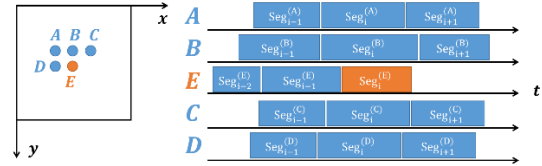

**Figure 8:** Spatiotemporal motion estimation.   **Figure 9:** Motion vector prediction for current segment.

The SME process is depicted in Fig. 8. In spatial, several previous coded pixels are selected as the reference pixels within a spatial range (SR). Meanwhile, in temporal, the reference segments are also determined by a temporal range (TR). By comparing the distance between current segment and each reference segment, the motion vector can be obtained which consists of $mv_X$, $mv_Y$ and $mv_T$ demonstrated in Fig. 8.

The motion vector prediction (MVP) is performed similar to that in forward mode. In inter mode, adjacent pixels are taken into account. Specifically, at most five pixels (A, B, C, D and E in Fig. 9) can be used for prediction. Here current segment $Seg_i^{(E)}$ in pixel E is to be coded, thus only its previous coded segments are available, such as $Seg_{i-1}^{(E)}$ and $Seg_{i-2}^{(E)}$. $i$ denotes the $i$-th segment in the pixel. For the other four pixels, three segments of each are utilized. One is the corresponding segment of $Seg_i^{(E)}$, including $Seg_i^{(A)}$, $Seg_i^{(B)}$, $Seg_i^{(C)}$ and $Seg_i^{(D)}$. The other two are the previous and the next segment in contrast to $Seg_i^{(E)}$. For example, in pixel A, $Seg_{i-1}^{(A)}$, $Seg_i^{(A)}$ and $Seg_{i+1}^{(A)}$ are available.

Thus, the average of the motion vectors of all the available reference segments is the MVP of $Seg_i^{(E)}$, including $mvPred_X$, $mvPred_Y$ and $mvPred_T$. Finally, the motion vector differences $mvd_X$, $mvd_Y$ and $mvd_T$ are encoded, where $mvd_X = mvPred_X - mv_X$, $mvd_Y = mvPred_Y - mv_Y$ and $mvd_T = mvPred_T - mv_T$. With the best reference segment as prediction, the residuals of ISIs are obtained and quantized.

### 4.4. Quantization

For spike data, the prediction residuals need to be quantized. Typically, the quantizer is designed to be a uniform one. However, according to Eq. (5), the same distortion of ISI may lead to completely different intensity changes. For instance, assuming that the distortion $\Delta t = 1$, the intensity change $\Delta I = \phi/2$ if $t = 1$, but when $t$ becomes 100, $\Delta I = \phi/10100$, which is much smaller than that when $t = 1$. Thus, a uniform quantization step is unfair for various ISIs.

To address the problem, the ISI itself is involved to the quantization, which means that each ISI should have a unique quantization step. One may question that how would decoder get the ISI value? Indeed, the raw ISI cannot be acquired by decoder, but owing to the well-designed prediction modes, it is instructive to utilize the predicted ISIs in the quantization. Specifically, in MVM, the mean value of the segment is available, so are the ISIs of the best reference segment in FM and inter-prediction mode. Eq. (8) describes the quantization, in which $R$ denotes the prediction residual, $C$ refers to the quantized residual, and $QStep(t)$ is the quantization step with the ISI of $t$. The function $round(\cdot)$ means rounding the residual $R$ to the nearest integer.

$$C = round\left(\frac{R}{Q_{step}(t)}\right) \tag{8}$$

Each quantization step $Q_{step}(t)$ at $t$ is defined as the maximum variation of ISI leading to neglected intensity change. Considering the function of rounding, the intensity keeps the same when $-0.5 < \Delta I < 0.5$. Thus, the maximum ISI variation $\Delta t$ follows $-\frac{t^2}{2\phi+t} < \Delta t < \frac{t^2}{2\phi-t}$. Due to $\left|\frac{t^2}{2\phi-t}\right| \geq \left|\frac{t^2}{2\phi+t}\right|$, by selecting the larger one, the quantization step is

$$Q_{step}(t) = \frac{t^2}{2\phi + t} \tag{9}$$

So the quantization step at each ISI can be computed according to Eq. (9). In this paper, we use a 5-bit (32 levels) quantizer.

## 5. Experimental Results

To evaluate the compression performance, the PKU-Spike dataset is constructed which contains six sequences including two categories of high speed motion and normal speed scenarios. Each sequence is captured by the spike camera at 40,000 Hz with a length of 3.84 seconds. As we discussed above, there are 32 levels with quantization parameter (QP) from 1 to 32. The spatial and temporal search range are set to 3 and 32, respectively. Table 1 demonstrates the compression ratio with various QPs in contrast to the raw data.

**Table 1:** Compression performance of proposed coding method using different QPs.

| Sequence | | Compression ratio (compared to the raw data) | | | | | | | |
|---|---|---|---|---|---|---|---|---|---|
| | | QP4 | QP8 | QP12 | QP16 | QP20 | QP24 | QP28 | QP32 |
| Normal speed | office | 67.21 | 78.38 | 82.99 | 90.54 | 103.87 | 115.10 | 127.86 | 141.08 |
| | rolling | 9.85 | 11.32 | 13.27 | 15.47 | 17.62 | 19.55 | 21.59 | 24.02 |
| | wavehand | 6.12 | 6.83 | 7.39 | 7.79 | 8.09 | 8.35 | 8.63 | 9.05 |
| High speed | fork | 43.29 | 58.52 | 74.62 | 91.74 | 105.76 | 115.44 | 122.63 | 128.91 |
| | disk-pku | 5.98 | 6.79 | 7.35 | 7.71 | 8.12 | 8.38 | 8.69 | 8.94 |
| | rotation | 5.76 | 6.41 | 6.93 | 7.32 | 7.66 | 7.97 | 8.17 | 8.48 |

Fig. 10 depicts the distortion caused by the compression. Both the intensity-based distance proposed in Section 3.2 and the kernel method based distance are utilized to measure the distortion. From the curves, the intensity-based distance is increasing along with the compression ratio, but in some sequences, such as "rolling" and "fork", the kernel method based distance may not clearly recognize the distortions.

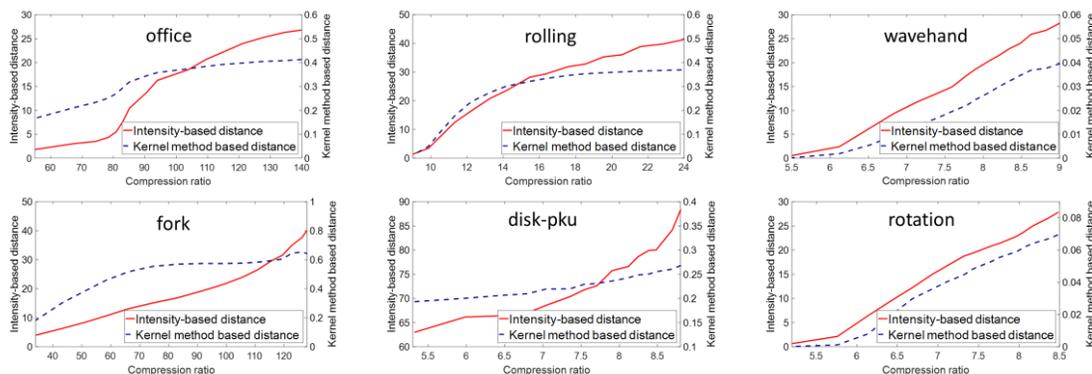

**Figure 10:** The distortion evaluation using the intensity-based distance and the kernel method based distance.

In addition, we reconstruct 1,000 images according to Eq. (4) from both the raw data and the decoded data. Then the images are evaluated via two commonly used metrics, the peak signal to noise ratio (PSNR) and the structural similarity index (SSIM), shown in Fig. 11. The result reveals that the fidelity of the spikes are well maintained.

## 6. Conclusion

In this paper, we aim at compressing the large amount of the spike data generated from the spike camera. In order to better evaluate the spike train distances, an intensity-based measurement is proposed according to the sampling mechanism. Then the spatiotemporal characteristics are deeply analyzed by modelling the inter-spike intervals. On the basis of the probability density distribution of ISIs, a unified lossy coding framework is designed comprising adaptive temporal partitioning, intra-/inter-pixel coding with multiple prediction modes, quantization and entropy coding. Finally, the experiment evaluations on the PKU-Spike dataset show the proposed coding method is quite efficient in compression while maintaining the fidelity of the spike data.

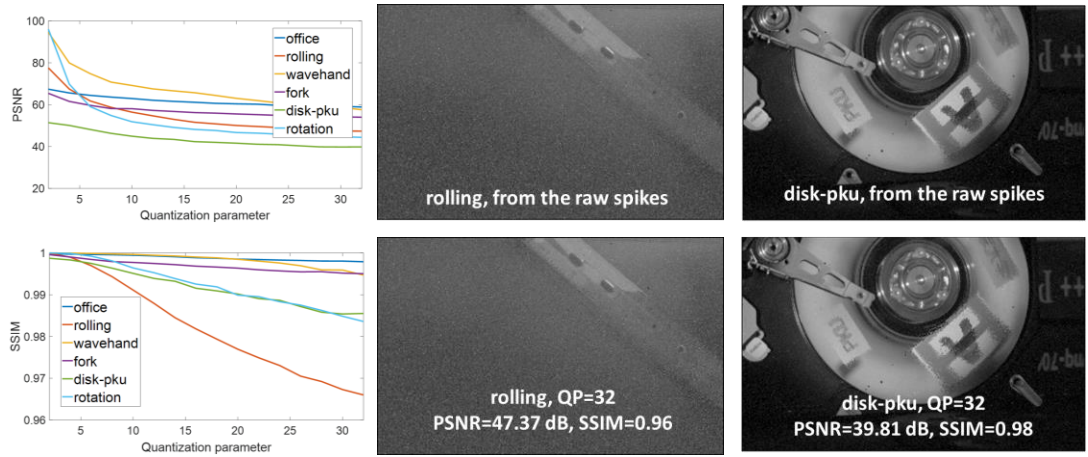

**Figure 11:** The measure of reconstructed images via PSNR and SSIM.